**The role of intangible investment in predicting stock returns:**

**Six decades of evidence**

Lin Li

Audencia Business School

Nantes, France

E-mail address: jack.li@audencia.com

I am grateful to Avanidhar Subrahmanyam, Zhaobo Zhu, Xin Chen, and an anonymous referee for providing useful comments. This research was conducted when I was at Shenzhen University. I gratefully acknowledge the research support from Shenzhen University (project number: 860-000002110719), Natural Science Foundation of Guangdong Province (project number: 2022A1515011952) and National Natural Science Foundation of China (project number: 72172095).

**Abstract**  Using an intangible intensity factor that is orthogonal to the Fama–French factors, we compare the role of intangible investment in predicting stock returns over the periods 1963–1992 and 1993–2022. For 1963–1992, intangible investment is weak in predicting stock returns, but for 1993–2022, the predictive power of intangible investment becomes very strong. Intangible investment has a significant impact not only on the MTB ratio (Fama-French HML factor) but also on operating profitability (Fama-French RMW factor) when forecasting stock returns from 1993 to 2022. For intangible asset-intensive firms, intangible investment is the main predictor of stock returns, rather than MTB ratio and profitability. Our evidence suggests that intangible investment has become an important factor in explaining stock returns over time, independent of other factors such as profitability and MTB ratio.

**Keywords:** Market-to-book ratio, intangible investment, profitability, stock returns, factor analysis

## 1. Introduction

Asset pricing has always been a hot topic in financial research, starting with the portfolio theory proposed by Markowitz (1952) and later the CAPM model pioneered by Sharpe (1964). Fama and French (1993) introduced size and value to explain stock returns on the basis of the CAPM model. Fama and French (2015) then added profitability and investment factors to their model to further capture stock returns. [1] Over the past three decades, the Fama-French framework has undergone many changes and evolutions as other researchers added their own factors and put their own spin on the duo's insights (e.g., Jegadeesh and Titman, 1993; Carhart, 1997; Pontiff and Schall, 1998; Hou et al., 2015).

Despite these efforts, few studies have explored the role of intangible investment in predicting stock returns. However, an indisputable fact is that U.S. companies have accelerated their shift from tangible to intangible investing over the past few decades, such as research and development (R&D) spending, the investment component of selling, general and administrative (SG&A) spending, and so on (Srivastava, 2014; Banker et al., 2019). Generally Accepted Accounting Principles (GAAP) require that R&D and SG&A spending be expensed as incurred. Since a large portion of intangible investments are immediately recorded in the income statement, company profitability and valuation as reflected by the MTB ratio are inevitably affected by this intangible investment accounting treatment (Enache and Srivastava, 2018), which in turn may affect their ability to explain stock returns. Therefore, intangible investment may have played a crucial role in determining company profitability and MTB ratios, as well as stock returns, over the past few decades.

---

[1] Note that the investment factor in Fama and French (2015) is constructed based on total assets and therefore primarily reflects investments in tangible assets.

This study investigates the role of intangible investment in predicting stock returns. Specifically, we want to know whether intangible investment has become an important factor in explaining stock returns and, if so, how the predictive power of other factors such as profitability and MTB ratio on stock returns is affected after accounting for intangible investment. Existing research suggests that intangible investment may be an important reason for the recent failure of MTB trading strategies (Arnott et al., 2021; Lev and Srivastava, 2022). Studies also find that the profitability factor can better explain stock returns after adjusting for intangible investment (Jagannathan et al., 2023; Rajgopal et al., 2024). However, it remains unknown whether intangible investment itself can be a factor in stock investment. This question is of great significance given the increase in intangible investment in the past few decades (Srivastava, 2014; Banker et al., 2019).

In this study, we compare the role of intangible investment in predicting stock returns over the periods 1963–1992 and 1993–2022. Following Fama and French (2015), we develop an intangible intensity factor based on R&D and the investment component of SG&A spending, and then test its predictive power within the framework of Fama-French five factors plus momentum. To mitigate the multicollinearity problem, we regress our constructed intangible intensity factor on Fama-French factors and use the orthogonal version of intangible intensity factor to predict stock returns.

Dividing the sample firms by MTB ratios, we find that intangible investment is weak in predicting stock returns over the 1963–1992 period, while the predictive power of the MTB ratio remains strong. Clearly, intangible investment has no significant impact on the ability of the MTB ratio to explain stock returns. In contrast, for 1993–2022, intangible investment becomes very strong in predicting stock returns, while the predictive power of MTB ratio on stock returns is largely suppressed by intangible investment. These findings demonstrate that intangible investment has become an important factor in explaining stock returns over the past decades. Moreover, intangible investing is an important factor that causes MTB trading strategy to lose its edge over time.

In addition to the MTB ratio, intangible investment is also highly correlated with the operating profitability between 1993 and 2022. Therefore, the ability of profitability to explain stock returns is also significantly affected, especially for less profitable firms. We further decompose the Fama-French profit factor into two parts, one part is related to intangible investment, and the other part is not related to intangible investment. For firms with low profitability rates, the predictive power of profitability on stock returns depends primarily on the intangible investment component. This evidence provides further support that intangible investment can be considered a predictor of stock returns. Furthermore, the function of intangible investment in explaining stock returns is not affected by the extrapolation bias identified by Lakonishok et al. (1994) and Bordalo et al. (2019).

To test the explanatory power of intangible investment on stock returns, we further divide the sample firms according to the intensity of their intangible investment. As for intangible itself, investors generally respond positively to corporate spending on intangible assets. When predicting stock returns from 1993 to 2022, intangible investment is the main predictor of stock returns for intangible asset-intensive firms, rather than MTB ratio and profitability. This evidence further confirms that intangible investment has become an important factor in explaining stock returns over time, independent of other factors such as profitability and MTB ratio.

We also find that the coefficients of market, size, and value factors in Fama-French five-factor model generally have larger magnitudes in 1963–1992 than in 1993–2022, implying that these three factors are better predictors of stock returns from 1963 to 1992.[2] In contrast, the coefficients of momentum and intangible investing are generally larger in absolute value for 1993–2022 than for 1963–1992, implying that momentum and intangible investing provide a stronger explanation for stock returns from 1993 to 2022. Clearly, the factors capturing stock returns are changing dynamically, and intangible investment becomes increasingly important in predicting stock returns over time.

Recent studies have documented that value strategy based on MTB ratio has lost its edge over time (Asness et al., 2015; McLean and Pontiff, 2016). We compare the role of intangible investment in predicting stock returns over two different periods, 1963–1992 and 1993–2022. The results show that the dysfunction of the MTB ratio in predicting stock returns occurs primarily in the most recent period, i.e., the period 1993–2022.[3] Moreover, we show that intangible investment has an increasingly significant impact on the MTB ratio's prediction of stock returns over time. Our findings complement existing research and help explain the recent weak performance of MTB trading strategies in the US market (Chen, 2017; Ball et al., 2018; Gerakos and Linnainmaa, 2018; Arnott et al., 2021; Lev and Srivastava, 2022; Goncalves and Leonard, 2023). Our findings also lend support to Hou et al. (2015), who construct a $q$-factor model consisting of market, size, investment, and profitability factors to explain the cross section of average stock return. While their model does a good job in explaining nearly 80 return anomalies, it fails to capture the R&D-to-Market anomaly. They suspect the failure is due to accounting rules requiring R&D spending to be expensed rather than capitalized.

---

[2] Smith and Timmermann (2022) document that the market equity, value, and size risk premia all vary significantly over time and have declined systematically over the nearly seven decades covered by their sample.

[3] Fama and French (2015) find that due to the inclusion of profitability and investment, the value factor (i.e., the MTB ratio) of the three-factor model becomes redundant in describing the average returns for a sample of U.S. firms from 1963 to 2013.

Moreover, our study demonstrates that intangible investment has as much, if not greater, impact on operating profitability (Fama-French RMW factor) as it does on MTB ratios (Fama-French HML factor) when predicting stock returns over the past few decades. This finding echoes Jagannathan et al. (2023), who show that profitability factors can better explain stock returns after adjusting for intangible investments. Rajgopal et al. (2024) show that capitalizing intangible assets can improve the assessment of firm profitability while increasing the usefulness of earnings measures. Of course, these evidences do not mean that intangible investment can completely absorb profitability or MTB ratio in explaining stock returns. [4] We, however, provide evidence that intangible investment is the key economic primitive behind operating profitability and MTB ratio, driving stock performance over the past few decades.

Importantly, unlike previous studies, our study directly demonstrates the role of intangible investment in predicting stock returns. Although the effect of intangible investment on the predictive power of profitability (Jagannathan et al., 2023; Rajgopal et al., 2024) or MTB ratio (Arnott et al., 2021; Lev and Srivastava, 2022) may somehow suggest that intangible investment contributes to stock returns, it remains unclear how the market prices intangible investment itself and whether intangible investment can be an independent factor in predicting stock returns that profitability and MTB ratio cannot substitute for. We show that, for companies with low profitability, the explanatory power of profitability on stock returns mainly comes from the intangible investment component; for companies with high intangible asset concentration, intangible investment is the main predictor of stock returns, not MTB ratio, tangible investment or operating profitability. These evidences clearly indicate that intangible investment can be considered an independent predictor of stock returns. We therefore call for special attention to intangible investing in portfolio management. This is particularly true given that both tech and non-tech companies in the U.S. have significantly increased their investments in intangible assets (Srivastava, 2014; Banker et al., 2019), especially as companies race to invest huge sums in R&D following the FinTech and artificial intelligence booms.

---

[4] Park (2019) shows that book-to-market ratio adjusted with R&D spending still has some explanatory power on stock performance.

Our study also sheds light on the issue of capitalization of intangible assets. Since GAAP requires that R&D and SG&A expenses be recorded as expenses when incurred, the actual profitability of enterprises is distorted to some extent, which reduces the quality of net income and price-to-earnings ratio in measuring enterprise value. To improve the value relevance of the income statement, accounting scholars have vigorously debated the issue of capitalization of intangible assets and presented a large amount of evidence for and against it (Lev and Sougiannis, 1996; Lev and Zarowin, 1999; Lev, 2008; Skinner, 2008; Zimmerman, 2015; Penman, 2023; Ball, 2024). This study directly tests whether and how intangible assets are priced by the market by constructing and exploring the role of intangible factors in stock return forecasting within the Fama-French framework. The results provide confidence that markets are efficient and capable of identifying the nature of intangible investments and thus pricing them correctly and adequately regardless of the form of financial reporting. Therefore, although the purpose of this study is not to participate in the debate on the capitalization of intangible assets, we believe that there is no need to worry too much about profitability bias under the current financial reporting system.

Our analysis proceeds as follows. Section 2 compares the role of intangible investment in predicting stock returns over the periods 1963–1992 and 1993–2022. Section 3 further examines the predictive power of intangible investment over the period 1993–2022. Section 4 concludes the paper.

## 2. Empirical Tests

### 2.1. Descriptive statistics

We start our analysis with descriptive statistics to understand the evolution of the MTB ratio and intangible investment over the past 60 years. Specifically, we look at MTB, ROE, R&D expenditures, SG&A expenditures, and intangible investments (INTAN) over the periods 1963–1992 and 1993–2022. INTAN is defined as R&D expenses plus the investment portion of selling, general & administrative (SG&A) expenses, divided by average total assets.[5] Data about R&D expense is obtained directly from the COMPUSTAT database, and observations with missing data are treated as having zero value. Observations with missing data to calculate MTB, ROE, and SG&A expenditures are deleted. Table 1 reports the values at the beginning, end, and entire period. In 1963, 1992, 1993 and 2022, the number of firms was 1,636, 6,065, 7,056 and 6,501 respectively.

(Insert Table 1 Here)

---

[5] Appendix A reports the detailed calculation of INTAN. The literature (Banker et al., 2011; Enache and Srivastava, 2018; Lev and Srivastava, 2022) document that a significant portion of SG&A expenses are related to intangible investments, such as in brands, human resources, and business processes. We refer to previous studies (Core et al., 1999; Roychowdhury, 2006; Enache and Srivastava, 2018) when calculating the investment component of SG&A expenses.

Overall, the average MTB ratio increases from 1.959 in 1963–1992 to 2.489 in 1993–2022, while ROE ratio decreases from 0.056 to -0.011. GAAP rules require R&D costs to be expensed in the year they were incurred, which will naturally reduce the net income for the year. Indeed, the average R&D expenditure as a percentage of revenue (RD) jumps from 1.9% in 1963–1992 to 5.1% in 1993–2022. SG&A expense is another factor that causes a low net income. The average SG&A expense as a percentage of revenue (SGA) jumps from 28.2% in 1963–1992 to 42.3% in 1993–2022. Consequently, the average intangible investments (INTAN) increase from 0.033 in 1963–1992 to 0.077 in 1993–2022. All changes are statistically significant.

Tech industry typically spends more on R&D than non-tech industries. Therefore, we also examine statistics for technology and non-tech firms separately. Following Chan et al. (2003), we classify technology firms using the following SIC codes: 283, 357, 366, 38, 48, or 737. From 1963–1992 to 1993–2022, R&D and SG&A expenditures increase in both technology and non-tech industries, but the growth is more pronounced in the technology industry. The average INTAN of technology firms jumps from 0.153 in 1963–1992 to 0.253 in 1993–2022, while the average INTAN of non-tech firms increases from 0.021 to 0.041.

Given high expenditures on intangible investments, the ROE ratio decreases, and the MTB ratio increases over time in both technology and non-tech industries, but this change is more obvious in the technology industry. For example, the average ROE of technology firms drops substantially from 0.013 in 1963–1992 to -0.109 in 1993–2022, while the average ROE of non-tech firms drops slightly from 0.060 to 0.011.

In summary, these results provide preliminary evidence that accounting earnings for U.S. firms decline as intangible investments increase, driving up the MTB ratio over time.

## 2.2. Construction of factors

From Wharton Research Data Services (WRDS) platform, we collect data on Fama-French factors, including market (MKTRF), size (SMB), book-to-market ratio (HML), profitability (RMW), and investment (CMA). We also collect data on momentum (UMD) factor from WRDS platform. Studies (e.g., Jegadeesh and Titman, 1993; Carhart, 1997) show that momentum contributes significantly to stock returns.

We develop an intangible intensity factor (INTANFT) to examine the role of intangible investment in predicting stock returns. Specifically, in June of each year t for the periods 1963–1992 and 1993–2022, all NYSE, Amex and NASDAQ stocks are divided into two groups, small and big (S and B), based on the NYSE median market cap. We then break these stocks into three intangible intensity groups for the bottom (Low), middle (Medium), and top (High) of the ranked values of INTAN. Therefore, the intangible intensity factor INTANFT is calculated as the difference, each month, between the simple average of the returns on the two high-INTAN portfolios (S/H and B/H) and the average of the returns on the two low-INTAN portfolios (S/L and B/L).

(Insert Table 2 Here)

Table 2 reports the Pearson correlations between INTANFT and other factors. Panel A reports correlations for the period 1963 to 1992, Panel B reports correlations for the period 1993 to 2022.

Several noteworthy results are illustrated. First of all, HML is significantly negatively related to the intangible intensity factor INTANFT in both periods. Specifically, the correlation coefficients of HML and INTANFT in 1963–1992 and 1993–2022 are -0.545 and -0.703 respectively. Obviously, HML has a stronger relationship with INTANFT from 1993 to 2022. Note that in Fama and French (1993, 2015) HML is calculated as the return on the high book-to-market (BTM) portfolio minus the return on the low BTM portfolio. Therefore, the negative correlation between INTANFT and Fama-French HML indicates that INTANFT is positively correlated with the MTB factor if we define HML as the return on the high MTB portfolio minus the return on the low MTB portfolio.

Second, INTANFT is significantly related to the profitability factor RMW in both periods. Specifically, the correlation coefficient of INTANFT and RMW in 1963–1992 and 1993–2022 are 0.237 and -0.742 respectively. Apparently, the association between INTANFT and RMW is more pronounced from 1993 to 2022. Note that the RMW factor in Fama and French (2015) is constructed based on operating profitability (OP), which is defined as revenue minus cost of sales, minus SG&A expenses, minus interest expense, divided by book equity. The increase in SG&A expenses during the period 1993–2022 naturally pushed up intangible investment while reducing OP, resulting in a negative correlation between INTANFT and RMW. This evidence is consistent with our previous finding that increases in intangible asset spending leads to declines in accounting performance over time.

Third, INTANFT is negatively related to the investment factor CMA for both periods. Specifically, the correlation coefficient of INTANFT and CMA in 1963–1992 and 1993–2022 are -0.438 and -0.495 respectively. Again, the association is more pronounced from 1993 to 2022. Note that the CMA in Fama and French (2015) is constructed based on total assets and therefore primarily reflects investments in tangible assets. Moreover, CMA is calculated as the return on the conservative investment portfolio minus the return on the aggressive investment portfolio.

Therefore, the negative correlation between INTANFT and Fama-French CMA indicates that INTANFT is positively correlated with the investment factor if we define CMA as the return on the aggressive investment portfolio minus the return on the conservative investment portfolio.

## 2.3. Univariate analysis

Before factor analysis, we first conduct a univariate analysis of monthly stock returns to gain an initial understanding of the impact of MTB ratio on stock returns. Specifically, in each year over the periods 1963–1992 and 1993–2022, we sort firms into 5 portfolios based on their MTB ratios at the end of the previous calendar year. We calculate monthly excess returns, R(t) – RF(t), on the 5 portfolios. R(t) is the average monthly returns on all the stocks in a portfolio. RF(t) is the one-month Treasury bill rate observed at the beginning of the month.

(Insert Table 3 Here)

Table 3 Panel A reports monthly excess returns for firms from 1963 to 1992 and 1993 to 2022, divided into five portfolios based on the MTB ratio. Between 1963 and 1992, the average monthly excess returns for low-MTB portfolios and high-MTB portfolios are 0.69% and 0.50%, respectively. Nonetheless, the difference between them is not statistically significant with a t-value of 0.39. Between 1993 and 2022, the low MTB portfolio has an average monthly return of 1.25%, while the high MTB portfolio has an average monthly return of 0.58%. The difference is also statistically insignificant, with a t-value of 1.37.

To examine the impact of intangible investment on stock returns, we next sort the firms by intangible investment INTAN at the end of the previous calendar year. Panel B of Table 3 reports the monthly excess returns for the five portfolios sorted by INTAN. Between 1963 and 1992, the average monthly return for the low-INTAN portfolio is 0.72%, which is smaller than the 0.91% for the high-INTAN portfolio (t-value = -0.40). Between 1993 and 2022, the average monthly return increases from 0.88% for the low INTAN portfolio to 1.05% for the high INTAN portfolio. But the increase is not statistically significant with a t-value of -0.32.

Overall, although the evidence suggests that the MTB ratio and intangible investment may affect stock returns, the difference in stock returns between the high and low ratio groups is not significant. Therefore, how they affect stock returns requires more rigorous factor analysis.

## 2.4.  Factor analysis

### 2.4.1.  MTB sorting

We now perform a more rigorous factor analysis of monthly excess stock returns with the following model:

$$R(t) - RF(t) = \alpha + b_1 * MKTRF(t) + b_2 * SMB(t) + b_3 * HML(t) + b_4 * RMW(t)$$
$$+ b_5 * CMA(t) + b_6 * UMD(t) + b_7 * INTANFT\ (t) + e(t) \qquad (1)$$

First, we sort our firms into 5 portfolios based on their MTB ratios at the end of the previous calendar year. The dependent variable in model (1), R(t) – RF(t), is the average monthly return on the portfolio of all sample stocks minus the one-month Treasury bill rate. [6] Table 4 reports the time series regression results for the periods 1963–1992 and 1993–2022. The period 1993–2022 covers 360 months while the period 1963–1992 covers 354 months, given that the Fama-French monthly factors in WRDS start from July 1963.

(Insert Table 4 Here)

Panel A reports regression results over the period 1963–1992. The key variables are HML and INTANFT. Look at HML first. For low MTB quintile, HML is significantly related to stock returns with an estimated coefficient of -0.270 (t-value=-3.23). For high MTB quintile, the coefficient of HML is -0.305 (t-value=-8.47). The magnitude of this coefficient is similar to those for low BTM quintiles (i.e., high MTB, ranging from -0.46 to -0.29) in Fama and French (1993), which analyze stock returns over the period 1963-1991 with a three-factor model. Importantly, all coefficients are statistically significant. This evidence suggests that MTB ratio has a significant impact on stock returns, even after controlling for other factors.

As for INTANFT, it is positively related to stock returns in all five regressions but maintains statistical significance at the 5% level or above only in quintiles with high MTB ratios. Specifically, INTANFT coefficient is 0.089 (t-value=1.99) in quintile 4 and 0.123 (t-value=2.66) in quintile 5 (i.e., high MTB). Apparently, although intangible investments have a positive impact on stock returns for high MTB firms, this effect is not evident for low MTB firms.

Panel B reports the regression results for the period 1993 to 2022. Generally, the magnitude of the HML coefficient decreases neatly from the low to the high MTB quintile, indicating that the MTB ratio has a significant impact on stock returns over the period. Importantly, INTANFT is significantly positive in all five regressions, suggesting that intangible investments have a strong impact on stock returns from 1993 to 2022. $R^2$ shows a growing trend from the low to the high MTB quintile between 1993 and 2022, just like 1963–1992.

However, if we compare the results between 1963–1992 and 1993–2022, two points catch our attention. First, for nearly all quintiles, the coefficients of MKTRF and SMB have larger magnitudes in 1963–1992 than in 1993–2022. This pattern also holds true for the absolute values of HML coefficients across most quintiles. For example, the HML coefficient for the highest MTB quintile during 1963–1992 is -0.305, but the HML coefficient for the highest MTB quintile during 1993–2022 is -0.140. This evidence suggests that the effect of the MTB ratio on stock returns weakens over time, which is consistent with previous research (Asness et al., 2015; McLean and Pontiff, 2016). This evidence also suggests that the three factors market, size, and MTB ratio are better at predicting stock returns from 1963 to 1992. Similarly, Smith and Timmermann (2022) also document that the market equity, value, and size risk premia all vary significantly over time and have declined systematically over the nearly seven decades covered by their sample.

In contrast, momentum and intangible investing provide a stronger explanation for stock returns from 1993 to 2022. For most quintiles, the coefficients of UMD and INTANFT are larger in absolute value for 1993–2022 than for 1963–1992. For example, in the low MTB quintile, UMD in 1993–2022 is statistically significant with an estimated coefficient of -0.396 and a t-value of -11.77, while the UMD coefficient in 1963–1992 is -0.014 and is not statistically significant (t-value =-0.37).

Second, for the five regressions in 1993–2022, four of them have statistically significant intercepts. Furthermore, for most quintiles, the magnitude of the intercept is larger for 1993–2022 than for 1963–1992. For example, the intercept for the low MTB quintile in 1993–2022 is 0.006 and the t-value is 3.72, but the intercept for the low-MTB quintile in 1963–1992 is 0.000 (t-value=0.24).

By contrast, for three of five quintiles, the $R^2$ value for 1993–2022 is smaller than the $R^2$ value for 1963–1992. For example, the $R^2$ for the low MTB quintile during 1993–2022 is 0.820, while the $R^2$ for the low MTB quintile during 1963–1992 is 0.875.

The combination of a larger intercept, smaller $R^2$, and smaller MKTRF coefficient (as mentioned above) indicates that stock prices provide more information about firm-specific characteristics. This evidence also indicates that there are more unknown risks affecting stock performance and therefore require more factors to predict during 1993–2022. [7] Morck et al. (2000) show that $R^2$ for U.S. firms trends downward over time when regressing the firm's own stock returns against market returns (i.e., one factor analysis). They argue that a lower $R^2$ means that a firm's stock return is less synchronous with the overall market and that a firm's stock price is more informative. Brogaard et al. (2022) show that since the mid-1990s, there has been a dramatic decline in noise and an increase in firm-specific information in stock price movements, consistent with increasing market efficiency.

The results in Table 4 suggest that the role of factors is changing dynamically, with intangible investments becoming increasingly important in predicting stock returns over time. One concern is that the intangible intensity factor INTANFT we constructed is highly correlated with other factors such as HML and RMW, especially for 1993–2022, which may lead to multicollinearity issues in the regression. To address this issue and examine the pure impact of intangible investments on stock returns, we regress INTANFT on Fama-French factors for 1993–2022:

$$\text{INTANFT} = 0.006 + 0.118*\text{MKTRF} + 0.221*\text{SMB} - 0.647*\text{HML}$$

$$(15.17) \quad (11.97) \qquad\qquad (15.35) \qquad (-39.36)$$

$$-0.633*\text{RMW} - 0.004*\text{CMA} + e \qquad\qquad\qquad (2)$$

$$(-35.21) \qquad\quad (-0.16)$$

---

[7] He and Zhou (2023) argue that the validity of asset pricing models implies white noise pricing errors, and therefore there is an urgent need to develop new asset pricing models. Kang and Baek (2024) show that debt heterogeneity can significantly affect equity returns.

The t-values are in parentheses below the slopes: the $R^2$ is 0.806. Similar to Fama and French (1993, 2015), the sum of the intercept and the residual in equation (2), call it INTANFT_Org, is used to predict stock returns during 1993–2022. Note that for the Pearson correlations of Table 2 Panel B, INTANFT is highly correlated with HML, RMW, and CMA with coefficients of -0.703, -0.742, and -0.495, respectively. In regression (2), the HML coefficient is -0.647 and the t-value is -39.36, and the RMW coefficient is -0.633 and the t-value is -35.21. In comparison, the CMA coefficient is only -0.004 and the t-value is only -0.16. Obviously, the association between CMA and INTANFT has been largely absorbed by HML and RMW.

Similarly, we also regress INTANFT on Fama-French factors for 1963–1992:

$$\text{INTANFT} = 0.001 + 0.078*\text{MKTRF} + 0.191*\text{SMB} - 0.329*\text{HML}$$

$$(4.54) \quad (12.01) \qquad\qquad (20.49) \qquad (-21.74)$$

$$+0.089*\text{RMW} + 0.133*\text{CMA} + e \qquad\qquad\qquad (3)$$

(4.25)          (6.10)

Again, we use the orthogonal version of intangible intensity factor INTANFT_Org to predict stock returns for the period 1963–1992. Note that the $R^2$ for equation (3) is 0.487, which is much smaller than 0.806 for equation (2), again confirming that INTANFT is more closely related to Fama-French factors during 1993–2022 than 1963–1992.

(Insert Table 5 Here)

We regress monthly excess stock returns on INTANFT_Org and other factors. Table 5 reports the results. The slopes on INTANFT_Org in all regressions in Table 5 are identical (by construction) to slopes on INTANFT in Table 4. Nevertheless, the slopes on Fama-French factors change.

In Panel A, for the period 1963 to 1992, the magnitude of the HML coefficient for each quintile decreases slightly relative to that in Table 4 Panel A. Importantly, the HML coefficient remains statistically significant across all quintiles. This evidence suggests that INTANFT has little effect on HML when predicting stock returns.

Panel B shows a different scenario. For each quintile, the decline in the HML coefficient from 1993 to 2022 is much greater than from 1963 to 1992. In the lowest MTB quintile, the HML coefficient drops from 0.319 in Panel B of Table 4 to 0.141. In the second highest MTB quintile, HML even becomes negative with coefficient -0.018, t-value -0.72, while in Panel B of Table 4 it is positive (coefficient = 0.043, t -value = 1.30). In the highest MTB quintile, HML is significantly negative with an estimated coefficient of -0.336 and a t-value of -11.64, while its coefficient is -0.140 with a t-value of -3.74 in Table 4 Panel B, indicating that the negative association between HML and stock returns is strengthened after orthogonalizing INTANFT. Apparently, the predictive power of the MTB ratio on stock returns is largely impacted by intangible investments.

Note that during 1993–2022, INTANFT is significantly positive in all regressions in Table 4 Panel B on the one hand and highly correlated with HML and RMW on the other hand, as shown by the Pearson correlations in Table 2 Panel B and regression model (2). Therefore, placing INTANFT together with HML and RMW in the model can definitely enhance the positive impact of the latter two on stock returns. By orthogonalizing INTANFT, its incremental positive impact on stock returns is also eliminated from HML and RMW. Consequently, the magnitude of coefficients for HML and RMW decreases sharply in the regressions.

In Panel B, the magnitude of the RMW coefficient also decreases sharply. For the low MTB quintile, RMW coefficient is -0.070 and bears no statistical significance (t-value = -0.81) in Table 4 Panel B, but its coefficient drops to -0.244 and becomes statistically significant (t-value = -3.53) in Table 5. For the high MTB quintile, the RMW coefficient in Table 5 Panel B is -0.352 (t-value = -11.52), while the RMW coefficient in Table 4 Panel B is -0.161 (t-value = -4.18). Clearly, the negative impact of RMW on stock returns is enhanced after eliminating the incremental positive impact of INTANFT. As we will further demonstrate later, while INTANFT is orthogonalized and the orthogonal version has no correlation with RMW, RMW itself contains a large portion of intangible investments, which largely determines its predictive power for stock returns.

In sum, the effect of the MTB ratio on stock returns weakens over time, while the effect of intangible investments increases. Over the past few decades, intangible investments have had a significant impact on the predictive power of the MTB ratio in predicting stock returns.

### 2.4.2. Intangible investment (INTAN) sorting

To clearly understand the impact of intangible investments on stock returns, we now sort our firms into 5 portfolios based on their INTAN ratios at the end of the previous calendar year. We calculate monthly excess returns, R(t) – RF(t), on 5 stock portfolios formed on INTAN. Table 6 Panel A reports the time series regression results of monthly excess returns on orthogonalized version of INTANFT, INTANFT_Org, and other factors for the 5 INTAN portfolios.

(Insert Table 6 Here)

Panel A1 reports regression results for the period 1963 to 1992. The magnitude of the INTANFT_Org coefficient generally increases from the low to the high INTAN quintile. By contrast, the magnitude of coefficients for both HML and RMW decreases with INTAN ratios. In particular, for high INTAN quintile, the coefficients for HML, RMW and INTANFT_Org are -0.190 (t-value= -4.19), -0.139 (t-value= -2.23) and 0.684 (t-value= 10.65) respectively. This evidence suggests that INTANFT_Org has a larger impact on stock returns relative to HML and RMW.

Panel A2 reports the regression results for the period 1993 to 2022. Again, the magnitude of the INTANFT_Org coefficients increases neatly from the low to the high INTAN quintile. For high INTAN quintiles, the absolute value of the coefficient of INTANFT_Org is larger than that of HML and RMW, indicating that INTANFT_Org has a larger impact on stock returns relative to RMW and HML.

Moreover, compared with 1963–1992, the growth trend of INTANFT_Org is more obvious from 1993 to 2022. For example, in the high INTAN quintile, the coefficient of INTANFT_Org reaches a striking 1.117, with a t-value of 16.66 in Panel A2, which is much larger and more significant than the 0.684 in Panel A1 (t-value = 10.65). The evidence suggests a stronger impact of intangible investments on stock returns during 1993–2022.

We also observe that the downward trend in the RMW and HML coefficients during 1993–2022 is more pronounced than during 1963–1992. For example, in Panel A2, for the high INTAN quintile, the RMW coefficient is -0.859 and the t-value is -15.36, which is much larger in absolute value and more significant than the -0.139 in Panel A1 (t-value = -2.23).

Overall, the impact of intangible investment on stock returns has increased over time. For intangible-intensive firms, intangible investment is a better predictor of stock returns than operating profitability and MTB ratios.

### 2.4.3. Operating profitability (OP) sorting

We next sort firms into 5 portfolios based on their operating profitability (OP) at the end of the previous calendar year. Following Fama and French (2015), OP is defined as revenue minus cost of sales, minus SG&A expenses, minus interest expense, divided by book equity. We calculate monthly excess returns, $R(t) - RF(t)$, on 5 stock portfolios formed on OP. Table 6 Panel

B reports the time series regression results of monthly excess returns on INTANFT_Org and other factors for the 5 OP portfolios.

We observe that for both periods, the RMW coefficient increases with increasing OP rate, while the INTANFT_Org coefficient decreases. But the pattern is more pronounced between 1993 and 2022. The evidence again confirms that operating profits and intangible investments have a greater impact on stock returns during 1993–2022 than during 1963–1992.

For firms with lower OP rates, the absolute value of the INTANFT_Org coefficient is almost as large as that of the RMW coefficient. For example, in the low OP quintile of Panel B2, the RMW coefficient is -0.968 and the t-value is -11.20, while the INTANFT_Org coefficient is 0.900 (t-value=8.68). This evidence suggests that in addition to earnings profitability, intangible investments also have a significant impact on stock returns for less profitable firms.

The SMB coefficients for the low OP quintile in Panel B1 and B2 are 1.425 and 0.881, respectively, which are much larger than the coefficients for the other quintiles. Not surprisingly, companies with low OP rates tend to be smaller, so the size factor explains more of stock returns.

The $R^2$ of the low OP quintiles in Panel B1 and B2 are 0.876 and 0.852 respectively, which are much smaller than the $R^2$ of other quintiles. This evidence suggests that more factors are needed to better explain the stock returns of firms with low OP rates.

## 2.5. Takeaways

The analysis in this section shows that the predictive power of intangible investment on stock returns increases over time. During 1993–2022, the correlation between intangible investment and the MTB ratio increases sharply. Likewise, intangible investment is also highly correlated with operating profitability. During this period, intangible investment has a strong power to predict stock returns, and when it is accounted for, the ability of MTB ratio and profitability to explain stock returns is significantly affected. For intangible-intensive companies, intangible investment is even the leading predictor of stock returns, rather than operating profitability and MTB ratio. All this evidence suggests that intangible investment has become an important factor in explaining stock returns over the past few decades, independent of other factors such as profitability and the MTB ratio. Even if intangible investment is not included as an independent factor in a model explaining stock returns, the HML and/or RMW factors should at least be adjusted to account for the impact of intangible investment. Jagannathan et al. (2023) and Rajgopal et al. (2024) find that the profitability factor better explains stock returns after adjusting for intangible investments. Cooper et al. (2024) show that the performance of the investment factors of Hou et al. (2015) and Fama and French (2015) depends largely on how their investment factors are constructed. If these measures take into account investments in intangible assets, their pricing power for stock returns is significantly reduced. Broman and Moneta (2024) find that fund managers tilt their portfolios toward the most salient factors in order to achieve high returns. Our results suggest that fund managers may also try to explore the role of intangible investments in their portfolio construction.

## 3. Further Analysis

Our research so far shows that intangible investments have a significant impact on MTB ratios and operating profitability when forecasting stock returns, particularly over the period 1993–2022. Given this fact, we now further analyze the role of intangible investments in predicting stock returns over the period.

## 3.1.  20 MTB–INTAN portfolios

We first divide the firms from 1993 to 2022 into 20 MTB-INTAN portfolios to examine the predictive power of intangible investments on stock returns. Firms are divided into five portfolios based on their MTB ratios at the end of the previous calendar year. Similarly, firms are also divided independently into four portfolios based on their INTAN ratios at the end of the previous calendar year. We perform time series regressions of monthly excess returns on INTANFT_Org and other factors for the 20 MTB–INTAN portfolios over the period 1993–2022. The results are presented in Table 7 Panel A. To save space, we report the intercepts, slopes for key variables HML, RMW, and INTANFT_Org, and $t$-statistics for these coefficients.

(Insert Table 7 Here)

Within each MTB quintile, the slope on INTANFT_Org increases monotonically from the low INTAN quartile to the high INTAN quartile. Of the 20 coefficients for INTANFT_Org (i.e., $b_7$), 16 are statistically significant above the 1% level. These results indicate that intangible investments have a significant impact on stock returns.

The impact is greater for intangible-intensive firms. For the 10 portfolios in the two quartiles with higher INTAN ratios, the INTANFT_Org coefficients are all significantly positive, with a minimum t-value of 5.89. In contrast, for the 10 HML coefficients (i.e., $b_3$), 3 are not statistically significant with t-values of -0.28, 0.45 and 0.69 respectively, and one is significant at the 5% level with t-value of -2.29. Furthermore, nearly all 10 INTANFT_Org coefficients are larger than the absolute values of the corresponding HML coefficients, regardless of MTB ranking. Clearly, the predictive power of the MTB ratio for stock returns is largely stifled by intangible investments.

The ability of operating profitability to explain stock returns is also significantly affected. Among the 20 portfolios, 4 RMW coefficients (i.e., $b_4$) are not statistically significant, and 5 are only significant at the 5% level. For the two quartiles with higher INTAN ratios, nearly all 10 INTANFT_Org coefficients are greater than the absolute values of the corresponding RMW coefficients.

Therefore, we are confident to conclude that intangible investment can affect MTB and profitability to a large extent when predicting stock returns, and even plays a leading role for intangible asset-intensive companies.

## 3.2.   20 OP–INTAN portfolios

We then divide the firms from 1993 to 2022 into 20 OP-INTAN portfolios. Specifically, firms are divided into five portfolios based on their OP rates at the end of the previous calendar year. Similarly, firms are also divided independently into four portfolios based on their INTAN ratios at the end of the previous calendar year. We perform time series regressions of monthly excess returns on INTANFT_Org and other factors for the 20 OP–INTAN portfolios over the period 1993–2022. Table 7 Panel B reports the intercepts, slopes for HML, RMW, and INTANFT_Org, and *t*-statistics for these coefficients.

Within each OP quintile, the slope on INTANFT_Org increases monotonically from the low to the high INTAN quartile. Among the 20 coefficients of INTANFT_Org (i.e., $b_7$), 14 are statistically significant above the 1% level, 2 are significant at the 5% level (t-value= -2.33 and -1.98). For the two quartiles with higher INTAN ratios, the INTANFT_Org coefficients are both positive and significant at the 1% level, and importantly, most coefficients are larger than the absolute values of the corresponding RMW coefficients (i.e., $b_4$).

Clearly, for intangible-intensive firms, intangible investments rather than operating profits play a dominant role in predicting stock returns. This is true even for more profitable firms. For example, for the portfolio with both high OP and high INTAN, the INTANFT_Org coefficient is 0.697 and the t-value is 8.67, which is larger and statistically more significant than the RMW coefficient of -0.288 and the t-value of -4.30.

## 3.3. Decomposing profitability factor RMW

Our research suggests that intangible investment has a substantial impact on operating profit when predicting stock returns over the period 1993 to 2022. According to the definition by Fama and French (2015), intangible investment such as SG&A expense is a significant component of operating profitability. Therefore, we decompose RMW into two parts by regressing RMW on INTANFT for the period 1993–2022: one related to intangible investment, RMW_INTAN, and the other not related to intangible investment, RMW_Org. Specifically, RMW_INTAN is defined as the predicted RMW of the regression, while RMW_Org is defined as the sum of the intercept and residual of the regression.

We divide firms into five portfolios based on their OP rates and regress monthly excess returns on the two parts of RMW while controlling for other factors for the 5 OP portfolios. Table 8 reports the results.

(Insert Table 8 Here)

Two notable results are illustrated. First, RMW_Org shows an increasing trend with OP rate and maintains statistical significance across all five quintiles. This evidence confirms the power of profitability in predicting stock returns, independent of intangible investments.

Second, RMW_INTAN predicts stock returns stronger for firms with lower OP rates. Specifically, in the lowest OP quintile, RMW_Org has a coefficient of -0.399 (t-value = -3.68), while RMW_INTAN has a coefficient as high as -2.2 with a t-value of -13.23. In the second lowest OP quintile, RMW_INTAN also has a higher coefficient of -1.020, which is statistically significant, with a t-value of -10.90, while RMW_Org only has a coefficient of -0.174 (t value = - 2.85). In addition, for the quintiles of middle-level and second-highest OP rates, the absolute value of RMW_INTAN is also larger than that of RMW_Org. Evidently, the predictive power of profitability on stock returns depends primarily on the intangible investment component, particularly for firms with lower OP rates. This evidence provides further support to our argument that intangible investment could be regarded as an independent factor in prediction stock returns.

3.4. Extrapolation bias

Finally, we divide the firms from 1993 to 2022 into 5 portfolios based on extrapolation bias in investor expectations. Lakonishok et al. (1994) show that extrapolation bias can significantly affect stock returns. La Porta (1996) and Bordalo et al. (2019) use analysts' expectations for a company's long-term earnings growth to proxy for extrapolation bias. They find that companies with optimistic earnings growth forecasts underperform compared to companies with pessimistic earnings growth forecasts.

Following these studies, we measure extrapolation bias using analysts' expectations for long-term earnings growth. We obtain mean analysts' forecasts for earnings per share (EPS) and their expected long-run growth rate (LTG) from IBES dataset. IBES defines LTG as the "expected annual increase in operating earnings over the company's next full business cycle", a period ranging from three to five years. On December of previous year between 1993 and 2022, we form LTG quintile portfolios for our firms. We perform time series regressions of monthly excess returns on INTANFT_Org and other factors for the 5 LTG quintile portfolios over the period 1993–2022. The results are presented in Table 9.

(Insert Table 9 Here)

Panel A reports the regression results for the entire sample period 1993–2022. We observe that INTANFT_Org is statistically significant in four of the five regressions. The magnitude of the coefficients increases neatly from the low to the high LTG quintile. In general, the absolute value of coefficient for INTANFT_Org is larger than that of momentum UMD. For portfolios with high LTG, the absolute value of coefficient for INTANFT_Org is also larger than that of HML and RMW. The evidence suggests a strong impact of intangible investments on stock returns during 1993–2022, regardless of extrapolation bias.

We also observe that the MKTRF and SMB coefficients increase with increasing LTG. One possible explanation is that analyst forecasts are more optimistic for smaller companies, which have larger betas and whose share prices are more likely to move with the market.

Particularly, we divide the sample into two groups according to the Internet bubble. Panel B reports the regression results for the Internet bubble period from 1995 to 2000, and Panel C reports the regression results for the non-Internet bubble period 1993–1994 and 2001–2022. For both panels, INTANFT_Org is statistically significant in most of five regressions, and the magnitude of the coefficients increases neatly from the low to the high LTG quintile. This evidence again suggests that intangible investments have a strong ability to predict stock returns.

In addition, by comparing the two panel data, we find that for investment portfolios with higher LTG, the absolute value of the INTANFT_Org coefficient during the Internet bubble period is smaller than that during the non-Internet bubble period, indicating that intangible investment has a stronger explanatory power for stock returns during the non-Internet bubble period; on the contrary, the absolute values of the UMD coefficients in the five regression models during the Internet bubble period are all greater than those during the non-Internet bubble period, which means that momentum contributes more to stock returns during the Internet bubble period.

Overall, the evidence confirms the important role of intangible investment in predicting stock returns, although this role may be moderated to some extent by investor enthusiasm.

## 4. Conclusion

We compare the role of intangible investment in predicting stock returns over the periods 1963–1992 and 1993–2022. For 1963–1992, intangible investment is weak in predicting stock returns, while the predictive power of the MTB ratio remains strong. In contrast, for 1993–2022, intangible investment becomes very strong in predicting stock returns, while the predictive power of MTB ratio on stock returns is largely suppressed by intangible investment.

Between 1993 and 2022, in addition to the MTB ratio, the ability of profitability to explain stock returns is also significantly affected by intangible investments, especially for less profitable firms. By decomposing the Fama-French profit factor, we find that for firms with low profitability, the ability of profitability to predict stock returns mainly depends on the intangible investment component, rather than components unrelated to intangible investment. For intangible asset-intensive firms, intangible investment is the main predictor of stock return, rather than MTB ratio and operating profit. Moreover, the explanation of intangible investment for stock returns is not affected by extrapolation bias in investor expectations.

Our study shows that intangible investment is the key economic primitive behind MTB ratio and accounting profitability that drives stock performance over the past few decades. Therefore, intangible investment could be regarded as an independent factor in explaining stock returns. We call for special attention to intangible investing when conducting portfolio management, given the increasing investment in intangible assets by U.S. businesses.

**Table 1**
**Descriptive statistics**

This table reports the descriptive statistics of MTB, ROE, R&D expenditures (RD), SG&A expenditures (SGA), and intangible investment (INTAN) over the periods 1963–1992 and 1993–2022. According to Chan et al. (2003), technology firms are classified as the following SIC codes: 283, 357, 366, 38, 48, or 737. We report both the mean and median (in parenthesis) value at the beginning, end, and entire period. In 1963, 1992, 1993 and 2022, the number of firms was 1,636, 6,065, 7,056 and 6,501 respectively. The definitions of variables are provided in Appendix A. For comparison of values between 1963–1992 and 1993–2022, the $t$-value and $z$-value are reported. *, **, and *** indicate significance at the 10%, 5%, and 1% levels, respectively.

| | 1963–1992 | | | 1993–2022 | | | $t$-value/$z$-value: |
|---|---|---|---|---|---|---|---|
| | 1963 | 1992 | 1963–1992 | 1993 | 2022 | 1993–2022 | (1963–1992 vs 1993–2022) |
| *All firms* | | | | | | | |
| MTB | 1.878 | 2.698 | 1.959 | 2.752 | 2.270 | 2.489 | 38.93*** |
| | (1.383) | (1.765) | (1.258) | (1.871) | (1.391) | (1.655) | 66.29*** |
| ROE | 0.107 | 0.010 | 0.056 | 0.023 | -0.096 | -0.011 | -26.21*** |
| | (0.105) | (0.074) | (0.108) | (0.083) | (0.054) | (0.070) | -64.73*** |
| RD | 0.004 | 0.037 | 0.019 | 0.034 | 0.066 | 0.051 | 85.87*** |
| | (0.000) | (0.000) | (0.000) | (0.000) | (0.000) | (0.000) | 51.13*** |
| SGA | 0.179 | 0.364 | 0.282 | 0.343 | 0.504 | 0.423 | 64.55*** |
| | (0.156) | (0.242) | (0.199) | (0.239) | (0.282) | (0.255) | 81.35*** |
| INTAN | 0.004 | 0.069 | 0.033 | 0.072 | 0.167 | 0.077 | 24.65*** |
| | (0.000) | (0.026) | (0.016) | (0.017) | (0.043) | (0.028) | 20.72*** |
| *Technology firms* | | | | | | | |
| MTB | 4.822 | 3.828 | 3.054 | 3.989 | 2.796 | 3.650 | 11.22*** |
| | (3.958) | (2.716) | (1.999) | (2.844) | (1.620) | (2.465) | 12.66*** |
| ROE | 0.155 | -0.039 | 0.013 | -0.037 | -0.336 | -0.109 | -12.96*** |
| | (0.150) | (0.083) | (0.108) | (0.065) | (-0.180) | (0.026) | -27.21*** |
| RD | 0.029 | 0.129 | 0.085 | 0.126 | 0.186 | 0.162 | 42.70*** |
| | (0.000) | (0.080) | (0.052) | (0.082) | (0.123) | (0.109) | 44.40*** |
| SGA | 0.302 | 0.553 | 0.444 | 0.559 | 0.845 | 0.721 | 39.89*** |
| | (0.301) | (0.419) | (0.347) | (0.418) | (0.581) | (0.503) | 51.55*** |
| INTAN | 0.059 | 0.237 | 0.153 | 0.263 | 0.446 | 0.253 | 12.56*** |
| | (0.085) | (0.226) | (0.128) | (0.214) | (0.387) | (0.206) | 12.15*** |

*Non-tech firms*

| | | | | | | | |
|---|---|---|---|---|---|---|---|
| MTB | 1.732 | 2.508 | 1.850 | 2.565 | 2.120 | 2.246 | 29.20*** |
| | (1.357) | (1.671) | (1.211) | (1.782) | (1.358) | (1.554) | 58.64*** |
| ROE | 0.105 | 0.018 | 0.060 | 0.033 | -0.027 | 0.011 | -19.13*** |
| | (0.103) | (0.072) | (0.108) | (0.085) | (0.077) | (0.075) | -54.92*** |
| RD | 0.003 | 0.023 | 0.012 | 0.020 | 0.036 | 0.028 | 50.05*** |
| | (0.000) | (0.000) | (0.000) | (0.000) | (0.000) | (0.000) | 12.81*** |
| SGA | 0.175 | 0.333 | 0.267 | 0.310 | 0.421 | 0.361 | 41.87*** |
| | (0.153) | (0.223) | (0.191) | (0.223) | (0.229) | (0.224) | 49.98*** |
| INTAN | 0.002 | 0.043 | 0.021 | 0.043 | 0.087 | 0.041 | 11.36*** |
| | (0.000) | (0.017) | (0.008) | (0.010) | (0.022) | (0.018) | 11.08*** |

**Table 2**
**Pearson correlations among factors**

This table reports the Pearson correlations among a set of factors. MKTRF (market return minus one-month Treasury bill rate) is the market factor; SMB (small minus big) is the size factor; HML (high minus low B/M) is the value factor; RMW (robust minus weak profitability) is the profitability factor; CMA (conservative minus aggressive investment) is the investment factor; UMD (up minus down return) is the momentum factor; and INTANFT (high minus low intangible investment) is the intangible intensity factor. Panel A reports correlations for the period 1963–1992. Panel B reports correlations for the period 1993–2022. Variable definitions are provided in Appendix A. The $p$-values are presented in the parenthesis. *, **, and *** denote statistical significance at the 10%, 5%, and 1% levels, respectively.

Panel A: Pearson correlations: 1963–1992

|         | MKTRF     | SMB       | HML       | RMW       | CMA       | UMD       | INTANFT |
|---------|-----------|-----------|-----------|-----------|-----------|-----------|---------|
| MKTRF   | 1         |           |           |           |           |           |         |
| SMB     | 0.338***  | 1         |           |           |           |           |         |
|         | (<.001)   |           |           |           |           |           |         |
| HML     | -0.361*** | -0.021    | 1         |           |           |           |         |
|         | (<.001)   | (0.329)   |           |           |           |           |         |
| RMW     | 0.076***  | -0.174*** | -0.537*** | 1         |           |           |         |
|         | (<.001)   | (<.001)   | (<.001)   |           |           |           |         |
| CMA     | -0.422*** | -0.172*** | 0.745***  | -0.553*** | 1         |           |         |
|         | (<.001)   | (<.001)   | (<.001)   | (<.001)   |           |           |         |
| UMD     | 0.022     | -0.119*** | -0.195*** | 0.201***  | -0.095*** | 1         |         |
|         | (0.307)   | (<.001)   | (<.001)   | (<.001)   | (<.001)   |           |         |
| INTANFT | 0.472***  | 0.400***  | -0.545*** | 0.237***  | -0.438*** | 0.086***  | 1       |
|         | (<.001)   | (<.001)   | (<.001)   | (<.001)   | (<.001)   | (<.001)   |         |

Panel B: Pearson correlations: 1993–2022

| | MKTRF | SMB | HML | RMW | CMA | UMD | INTANFT |
|---|---|---|---|---|---|---|---|
| MKTRF | 1 | | | | | | |
| SMB | 0.238*** | 1 | | | | | |
| | (<.001) | | | | | | |
| HML | -0.086*** | -0.031 | 1 | | | | |
| | (<.001) | (0.156) | | | | | |
| RMW | -0.346*** | -0.458*** | 0.364*** | 1 | | | |
| | (<.001) | (<.001) | (<.001) | | | | |
| CMA | -0.311*** | -0.035 | 0.645*** | 0.246*** | 1 | | |
| | (<.001) | (0.105) | (<.001) | (<.001) | | | |
| UMD | -0.306*** | -0.025 | -0.212*** | 0.035 | 0.024 | 1 | |
| | (<.001) | (0.240) | (<.001) | (0.106) | (0.263) | | |
| INTANFT | 0.364*** | 0.409*** | -0.703*** | -0.742*** | -0.495*** | 0.037* | 1 |
| | (<.001) | (<.001) | (<.001) | (<.001) | (<.001) | (0.088) | |

**Table 3**
**Monthly excess stock returns sorted by MTB and intangible investments**

This table reports the monthly excess stock returns, R(t) − RF(t), for firms sorted to 5 portfolios by MTB ratios or intangible investing INTAN at the end of the previous calendar year over the periods 1963–1992 and 1993–2022. R(t) is average monthly returns on all the stocks in a portfolio. RF(t) is the one-month Treasury bill rate observed at the beginning of the month. Variable definitions are provided in Appendix A. We report both the mean and median value for monthly stock returns. The *t*-values are presented in the parenthesis to test the mean value. For comparison of stock returns, the *t*-value and *z*-value are reported. *, **, and *** denote statistical significance at the 10%, 5%, and 1% levels, respectively.

Panel A: Monthly excess returns for 5 portfolios formed on MTB ratios

|  | 1963–1992 | | 1993–2022 | |
|---|---|---|---|---|
|  | Average | Median | Average | Median |
| Low MTB | 0.0069** | 0.0066 | 0.0125*** | 0.0146 |
|  | (2.01) |  | (3.64) |  |
| 2 | 0.0099*** | 0.0083 | 0.0103*** | 0.0135 |
|  | (3.24) |  | (3.51) |  |
| 3 | 0.0091*** | 0.0099 | 0.0100*** | 0.0143 |
|  | (2.72) |  | (3.62) |  |
| 4 | 0.0065** | 0.0091 | 0.0083*** | 0.0134 |
|  | (2.12) |  | (2.86) |  |
| High MTB | 0.0050 | 0.0093 | 0.0058* | 0.0095 |
|  | (1.49) |  | (1.66) |  |
| Low vs High |  |  |  |  |
| (*t*-value/*z*-value) | 0.39 | 0.17 | 1.37 | 0.89 |

Panel B: Monthly excess returns for 5 portfolios formed on INTAN ratios

|  | 1963–1992 | | 1993–2022 | |
|---|---|---|---|---|
|  | Average | Median | Average | Median |
| Low INTAN | 0.0072** | 0.0071 | 0.0088*** | 0.0130 |
|  | (2.43) |  | (3.26) |  |
| 2 | 0.0078** | 0.0082 | 0.0093*** | 0.0119 |
|  | (2.48) |  | (3.44) |  |
| 3 | 0.0081** | 0.0093 | 0.0097*** | 0.0139 |
|  | (2.44) |  | (3.01) |  |
| 4 | 0.0085** | 0.0067 | 0.0111*** | 0.0133 |
|  | (2.44) |  | (3.01) |  |
| High INTAN | 0.0091** | 0.0061 | 0.0105** | 0.0117 |
|  | (2.53) |  | (2.27) |  |
| Low vs High |  |  |  |  |
| (*t*-value/*z*-value) | -0.40 | -0.13 | -0.32 | -0.20 |

**Table 4**
**Regressing monthly excess returns on intangible intensity factor by MTB sorting**

This table reports the regression results of monthly excess returns, R(t) − RF(t), on a set of factors with the following model:

$$R(t) - RF(t) = \alpha + b_1 * MKTRF(t) + b_2 * SMB(t) + b_3 * HML(t) + b_4 * RMW(t) + b_5 * CMA(t) + b_6 * UMD(t) + b_7 * INTANFT(t) + e(t)$$

Firms are divided into five portfolios based on their MTB ratios at the end of the previous calendar year. R(t) is the average monthly returns on all the stocks in a portfolio. RF(t) is the one-month Treasury bill rate observed at the beginning of the month. MKTRF (market return minus one-month Treasury bill rate) is the market factor; SMB (small minus big) is the size factor; HML (high minus low B/M) is the value factor; RMW (robust minus weak profitability) is the profitability factor; CMA (conservative minus aggressive investment) is the investment factor; UMD (up minus down return) is the momentum factor; and INTANFT (high minus low intangible investment) is the intangible intensity factor. Panel A reports regression results for the period 1963−1992, 354 months. Panel B reports regression results for the period 1993−2022, 360 months. Variable definitions are provided in Appendix A. The *t*-values are presented in parenthesis. *, **, and *** denote statistical significance at the 10%, 5%, and 1% levels, respectively.

Panel A: Regressing monthly excess returns on factors in 1963−1992

|          | intercept | MKTRF | SMB | HML | RMW | CMA | UMD | INTANFT | $R^2$ |
|----------|-----------|-------|-----|-----|-----|-----|-----|---------|-------|
| Low MTB  | 0.000     | 0.915*** | 1.069*** | -0.270*** | -0.252** | 0.380*** | -0.014 | 0.115 | 0.875 |
|          | (0.24)    | (27.51) | (21.09) | (-3.23) | (-2.41) | (3.49) | (-0.37) | (1.07) | |
| 2        | 0.003**   | 0.851*** | 0.953*** | 0.338*** | -0.045 | 0.079 | -0.135*** | 0.160* | 0.908 |
|          | (2.63)    | (33.63) | (24.72) | (5.32) | (-0.57) | (0.95) | (-4.77) | (1.95) | |
| 3        | 0.001     | 0.951*** | 1.042*** | 0.546*** | 0.020 | -0.077 | -0.152*** | 0.124 | 0.919 |
|          | (0.89)    | (36.37) | (26.16) | (8.32) | (0.24) | (-0.90) | (-5.23) | (1.46) | |
| 4        | 0.000     | 0.986*** | 0.791*** | 0.205*** | -0.138*** | -0.049 | -0.106*** | 0.089** | 0.973 |
|          | (0.69)    | (70.77) | (37.23) | (5.87) | (-3.15) | (-1.07) | (-6.82) | (1.99) | |
| High MTB | 0.001     | 1.026*** | 0.749*** | -0.305*** | -0.284*** | -0.142*** | -0.025 | 0.123*** | 0.976 |
|          | (1.57)    | (71.41) | (34.18) | (-8.47) | (-6.28) | (-3.02) | (-1.54) | (2.66) | |

Panel B: Regressing monthly excess returns on factors in 1993–2022

| | intercept | MKTRF | SMB | HML | RMW | CMA | UMD | INTANFT | $R^2$ |
|---|---|---|---|---|---|---|---|---|---|
| Low MTB | 0.006*** | 0.867*** | 0.561*** | 0.319*** | -0.070 | 0.147 | -0.396*** | 0.275*** | 0.820 |
| | (3.72) | (21.50) | (9.62) | (3.75) | (-0.81) | (1.54) | (-11.77) | (3.32) | |
| 2 | 0.004*** | 0.831*** | 0.655*** | 0.339*** | 0.026 | 0.015 | -0.281*** | 0.108** | 0.922 |
| | (4.39) | (36.75) | (20.00) | (7.11) | (0.53) | (0.27) | (-14.86) | (2.32) | |
| 3 | 0.003*** | 0.905*** | 0.657*** | 0.223*** | 0.118*** | 0.082** | -0.153*** | 0.064* | 0.950 |
| | (3.74) | (52.92) | (26.53) | (6.19) | (3.21) | (2.04) | (-10.76) | (1.83) | |
| 4 | 0.001 | 0.975*** | 0.654*** | 0.043 | 0.108*** | 0.040 | -0.113*** | 0.094*** | 0.962 |
| | (1.02) | (62.66) | (29.03) | (1.30) | (3.23) | (1.09) | (-8.68) | (2.93) | |
| High MTB | -0.002** | 1.030*** | 0.645*** | -0.140*** | -0.161*** | -0.059 | -0.033** | 0.303*** | 0.967 |
| | (-2.32) | (57.78) | (25.00) | (-3.74) | (-4.18) | (-1.41) | (-2.21) | (8.27) | |

**Table 5**

**Regressing monthly excess returns on orthogonal version of intangible intensity factor by MTB sorting**

This table reports the regression results of monthly excess returns, R(t) − RF(t), on a set of factors with the following model:

$$R(t) - RF(t) = \alpha + b_1 * MKTRF(t) + b_2 * SMB(t) + b_3 * HML(t) + b_4 * RMW(t) + b_5 * CMA(t) + b_6 * UMD(t) + b_7 * INTANFT\_Org(t) + e(t)$$

Panel A reports regression results for the period 1963−1992, 354 months. Panel B reports regression results for the period 1993−2022, 360 months. Firms are divided into five portfolios based on their MTB ratios at the end of the previous calendar year. R(t) is the average monthly returns on all the stocks in a portfolio. RF(t) is the one-month Treasury bill rate observed at the beginning of the month. MKTRF (market return minus one-month Treasury bill rate) is the market factor; SMB (small minus big) is the size factor; HML (high minus low B/M) is the value factor; RMW(robust minus weak profitability) is the profitability factor; CMA (conservative minus aggressive investment) is the investment factor; UMD (up minus down return) is the momentum factor; and INTANFT_Org is orthogonal version of intangible intensity factor INTANFT (high minus low intangible investment), calculated as the sum of the intercept and the residual in the following regression run separately for the periods 1963−1992 and 1993−2022: INTANFT = a + b*MKTRF + c*SMB +d*HML+e*RMW+ f*CMA +e. Variable definitions are provided in Appendix A. The t-values are presented in the parenthesis. *, **, and *** denote statistical significance at the 10%, 5%, and 1% levels, respectively.

Panel A: Regressing monthly excess returns on factors in 1963−1992

| | intercept | MKTRF | SMB | HML | RMW | CMA | UMD | INTANFT_Org | $R^2$ |
|---|---|---|---|---|---|---|---|---|---|
| Low MTB | 0.000 | 0.924*** | 1.091*** | -0.307*** | -0.242** | 0.396*** | -0.014 | 0.115 | 0.875 |
| | (0.24) | (28.71) | (23.57) | (-4.06) | (-2.32) | (3.66) | (-0.37) | (1.07) | |
| 2 | 0.003*** | 0.863*** | 0.984*** | 0.285*** | -0.031 | 0.100 | -0.135*** | 0.160* | 0.908 |
| | (2.63) | (35.27) | (27.94) | (4.95) | (-0.39) | (1.22) | (-4.77) | (1.95) | |
| 3 | 0.001 | 0.960*** | 1.066*** | 0.505*** | 0.031 | -0.061 | -0.152*** | 0.124 | 0.919 |
| | (0.89) | (37.96) | (29.29) | (8.49) | (0.38) | (-0.71) | (-5.23) | (1.46) | |
| 4 | 0.000 | 0.993*** | 0.808*** | 0.176*** | -0.130*** | -0.037 | -0.106*** | 0.089** | 0.973 |
| | (0.69) | (73.65) | (41.64) | (5.55) | (-2.98) | (-0.81) | (-6.82) | (1.99) | |
| High MTB | 0.001 | 1.036*** | 0.772*** | -0.346*** | -0.273*** | -0.126*** | -0.025 | 0.123*** | 0.976 |
| | (1.57) | (74.49) | (38.60) | (-10.58) | (-6.06) | (-2.69) | (-1.54) | (2.66) | |

Panel B: Regressing monthly excess returns on factors in 1993–2022

| | intercept | MKTRF | SMB | HML | RMW | CMA | UMD | INTANFT_Org | $R^2$ |
|---|---|---|---|---|---|---|---|---|---|
| Low MTB | 0.006*** | 0.899*** | 0.622*** | 0.141** | -0.244*** | 0.146 | -0.396*** | 0.275*** | 0.820 |
| | (3.72) | (22.92) | (11.23) | (2.16) | (-3.53) | (1.53) | (-11.77) | (3.32) | |
| 2 | 0.004*** | 0.844*** | 0.678*** | 0.269*** | -0.042 | 0.014 | -0.280*** | 0.108** | 0.922 |
| | (4.39) | (38.35) | (21.85) | (7.34) | (-1.09) | (0.27) | (-14.86) | (2.32) | |
| 3 | 0.003*** | 0.913*** | 0.671*** | 0.182*** | 0.077*** | 0.082** | -0.153*** | 0.064* | 0.950 |
| | (3.74) | (54.85) | (28.56) | (6.54) | (2.64) | (2.03) | (-10.76) | (1.83) | |
| 4 | 0.001 | 0.986*** | 0.674*** | -0.018 | 0.049* | 0.039 | -0.113*** | 0.094*** | 0.962 |
| | (1.02) | (65.14) | (31.56) | (-0.72) | (1.82) | (1.08) | (-8.68) | (2.93) | |
| High MTB | -0.002** | 1.065*** | 0.712*** | -0.336*** | -0.352*** | -0.060 | -0.033** | 0.303*** | 0.966 |
| | (-2.32) | (61.45) | (29.08) | (-11.64) | (-11.52) | (-1.44) | (-2.21) | (8.27) | |

**Table 6**

**Regressing monthly excess returns on orthogonal version of intangible intensity factor by INTAN or OP sorting**

This table reports the regression results of monthly excess returns, $R(t) - RF(t)$, on a set of factors over the periods 1963–1992 (354 months) and 1993–2022 (360 months). The model is as follows:

$$R(t) - RF(t) = \alpha + b_1 * MKTRF(t) + b_2 * SMB(t) + b_3 * HML(t) + b_4 * RMW(t) + b_5 * CMA(t) + b_6 * UMD(t) + b_7 * INTANFT\_Org\,(t) + e(t)$$

Panel A reports the regression results for 5 INTAN portfolios. Firms are divided into five portfolios based on their INTAN ratios at the end of the previous calendar year, calculated as R&D expenses plus the investment portion of selling, general & administrative (SG&A) expenses, divided by average total assets. Panel B reports the regression results for 5 OP portfolios. Firms are divided into five portfolios based on their operating profitability (OP) at the end of the previous calendar year, defined as revenue minus cost of sales, minus SG&A expenses, minus interest expense, divided by book equity. $R(t)$ is the average monthly returns on all the stocks in a portfolio. $RF(t)$ is the one-month Treasury bill rate observed at the beginning of the month. MKTRF (market return minus one-month Treasury bill rate) is the market factor; SMB (small minus big) is the size factor; HML (high minus low B/M) is the value factor; RMW(robust minus weak profitability) is the profitability factor; CMA (conservative minus aggressive investment) is the investment factor; UMD (up minus down return) is the momentum factor; and INTANFT_Org is orthogonal version of intangible intensity factor INTANFT (high minus low intangible investment), calculated as the sum of the intercept and the residual in the following regression run separately for the periods 1963–1992 and 1993–2022: $INTANFT = a + b*MKTRF + c*SMB + d*HML + e*RMW + f*CMA + e$. Variable definitions are provided in Appendix A. The *t*-values are presented in the parenthesis. *, **, and *** denote statistical significance at the 10%, 5%, and 1% levels, respectively.

Panel A: Regressing monthly excess returns on factors for 5 INTAN portfolios

A1: Regression results for 1963−1992

| | intercept | MKTRF | SMB | HML | RMW | CMA | UMD | INTANFT_Org | $R^2$ |
|---|---|---|---|---|---|---|---|---|---|
| Low INTAN | 0.001** | 0.924*** | 0.860*** | 0.215*** | -0.127*** | -0.033 | -0.077*** | -0.251*** | 0.975 |
| | (2.21) | (74.51) | (48.20) | (7.39) | (-3.15) | (-0.79) | (-5.38) | (-6.07) | |
| 2 | 0.001 | 0.959*** | 0.930*** | 0.170*** | -0.142 | 0.026 | -0.054*** | 0.006 | 0.965 |
| | (1.39) | (62.18) | (41.89) | (4.68) | (-2.85) | (0.49) | (-3.06) | (0.12) | |
| 3 | 0.001 | 0.991*** | 1.019*** | 0.033 | -0.140*** | 0.091 | -0.065*** | 0.356*** | 0.971 |
| | (1.34) | (66.66) | (47.65) | (0.95) | (-2.90) | (1.82) | (-3.82) | (7.18) | |
| 4 | 0.001 | 0.996*** | 1.093*** | -0.025 | -0.086 | 0.024 | -0.051*** | 0.586*** | 0.969 |
| | (1.36) | (61.92) | (47.28) | (-0.67) | (-1.65) | (0.44) | (-2.76) | (10.92) | |
| High INTAN | 0.002** | 0.965*** | 1.171*** | -0.190*** | -0.139** | 0.045 | -0.055** | 0.684*** | 0.959 |
| | (2.48) | (50.11) | (42.27) | (-4.19) | (-2.23) | (0.70) | (-2.48) | (10.65) | |

A2: Regression results for 1993−2022

| | intercept | MKTRF | SMB | HML | RMW | CMA | UMD | INTANFT_Org | $R^2$ |
|---|---|---|---|---|---|---|---|---|---|
| Low INTAN | 0.003*** | 0.901*** | 0.569*** | 0.288*** | 0.104*** | 0.038 | -0.163*** | -0.238*** | 0.947 |
| | (4.70) | (53.94) | (24.13) | (10.33) | (3.52) | (0.93) | (-11.36) | (-6.75) | |
| 2 | 0.003*** | 0.897*** | 0.600*** | 0.278*** | 0.143*** | 0.041 | -0.170*** | -0.087** | 0.938 |
| | (3.68) | (49.46) | (23.43) | (9.19) | (4.47) | (0.94) | (-10.95) | (-2.27) | |
| 3 | 0.001 | 1.022*** | 0.700*** | -0.056 | -0.115*** | 0.111** | -0.197*** | 0.347*** | 0.944 |
| | (1.23) | (49.74) | (24.13) | (-1.64) | (-3.17) | (2.23) | (-11.20) | (7.99) | |
| 4 | 0.002** | 1.016*** | 0.802*** | -0.332*** | -0.431*** | 0.094* | -0.203*** | 0.711*** | 0.952 |
| | (2.12) | (46.55) | (26.01) | (-9.14) | (-11.22) | (1.77) | (-10.82) | (15.42) | |
| High INTAN | 0.000 | 1.040*** | 0.967*** | -0.554*** | -0.859*** | 0.036 | -0.194*** | 1.117*** | 0.934 |
| | (0.35) | (32.76) | (21.58) | (-10.47) | (-15.36) | (0.47) | (-8.58) | (16.66) | |

Panel B: Regressing monthly excess returns on factors for 5 OP portfolios

**B1: Regression results for 1963–1992**

|          | intercept | MKTRF | SMB | HML | RMW | CMA | UMD | INTANFT_Org | $R^2$ |
|----------|-----------|-------|-----|-----|-----|-----|-----|-------------|-------|
| Low OP   | -0.000    | 0.934*** | 1.425*** | 0.058 | -0.528*** | 0.056 | -0.055 | 0.444*** | 0.876 |
|          | (-0.20)   | (25.15) | (26.69) | (0.66) | (-4.39) | (0.45) | (-1.29) | (3.58) | |
| 2        | 0.000     | 0.918*** | 1.055*** | 0.071* | -0.270*** | 0.073 | -0.083*** | 0.291*** | 0.958 |
|          | (0.56)    | (52.14) | (41.64) | (1.70) | (-4.73) | (1.23) | (-4.09) | (4.96) | |
| 3        | 0.001*    | 0.966*** | 0.906*** | 0.041 | -0.070** | 0.092** | -0.100*** | 0.168*** | 0.982 |
|          | (1.77)    | (88.43) | (57.61) | (1.60) | (-1.97) | (2.51) | (-7.98) | (4.59) | |
| 4        | 0.001***  | 0.996*** | 0.873*** | 0.056** | 0.141*** | -0.010 | -0.050*** | 0.176*** | 0.986 |
|          | (2.73)    | (102.13) | (62.28) | (2.45) | (4.47) | (-0.29) | (-4.46) | (5.40) | |
| High OP  | 0.002***  | 1.041*** | 1.079*** | -0.093** | 0.058 | 0.060 | -0.014 | 0.197*** | 0.965 |
|          | (2.70)    | (59.55) | (42.93) | (-2.26) | (1.03) | (1.02) | (-0.69) | (3.38) | |

**B2: Regression results for 1993–2022**

|          | intercept | MKTRF | SMB | HML | RMW | CMA | UMD | INTANFT_Org | $R^2$ |
|----------|-----------|-------|-----|-----|-----|-----|-----|-------------|-------|
| Low OP   | -0.001    | 1.008*** | 0.881*** | -0.365*** | -0.968*** | 0.062 | -0.355*** | 0.900*** | 0.852 |
|          | (-0.25)   | (20.54) | (12.71) | (-4.46) | (-11.20) | (0.52) | (-8.44) | (8.68) | |
| 2        | 0.001     | 0.902*** | 0.699*** | -0.160*** | -0.441*** | 0.124* | -0.250*** | 0.423*** | 0.907 |
|          | (1.14)    | (32.65) | (17.93) | (-3.47) | (-9.07) | (1.85) | (-10.57) | (7.25) | |
| 3        | 0.002***  | 0.937*** | 0.724*** | 0.102*** | 0.022 | 0.042 | -0.154*** | 0.106*** | 0.946 |
|          | (3.15)    | (51.60) | (28.24) | (3.36) | (0.68) | (0.95) | (-9.88) | (2.76) | |
| 4        | 0.002***  | 0.980*** | 0.655*** | 0.146*** | 0.285*** | 0.013 | -0.141*** | -0.039 | 0.962 |
|          | (4.05)    | (67.64) | (32.04) | (6.04) | (11.18) | (0.37) | (-11.32) | (-1.28) | |
| High OP  | 0.002***  | 1.051*** | 0.644*** | 0.045 | 0.189*** | 0.014 | -0.164*** | 0.094** | 0.937 |
|          | (2.70)    | (51.64) | (22.41) | (1.32) | (5.27) | (0.29) | (-9.40) | (2.19) | |

**Table 7**
**Regressing monthly excess returns on orthogonal version of intangible intensity factor by OP, MTB, and INTAN sorting**

**during 1993–2022**

Monthly excess returns, R(t) – RF(t), is regressed on a set of factors over the period 1993–2022 (360 months) with the following model:

R(t) – RF(t) = α + $b_1$ * MKTRF(t) + $b_2$ * SMB(t) + $b_3$ * HML(t) + $b_4$ * RMW(t) + $b_5$ * CMA(t) + $b_6$ * UMD(t) + $b_7$ * INTANFT_Org (t) + $e$(t)

Panel A reports the regression results for 20 MTB-INTAN portfolios. Firms are divided into five portfolios based on their MTB ratios at the end of the previous calendar year. Similarly, firms are also divided independently into four portfolios based on their INTAN ratios at the end of the previous calendar year, calculated as R&D expenses plus the investment portion of selling, general & administrative (SG&A) expenses, divided by average total assets. Panel B reports the regression results for 20 OP-INTAN portfolios. Firms are divided into five portfolios based on their operating profitability (OP) at the end of the previous calendar year, defined as revenue minus cost of sales, minus SG&A expenses, minus interest expense, divided by book equity. Firms are also divided independently into four portfolios based on their INTAN ratios at the end of the previous calendar year. R(t) is the average monthly returns on all the stocks in a portfolio. RF(t) is the one-month Treasury bill rate observed at the beginning of the month. MKTRF (market return minus one-month Treasury bill rate) is the market factor; SMB (small minus big) is the size factor; HML (high minus low B/M) is the value factor; RMW(robust minus weak OP) is the profitability factor; CMA (conservative minus aggressive investment) is the investment factor; UMD (up minus down return) is the momentum factor; and INTANFT_Org is orthogonal version of intangible intensity factor INTANFT (high minus low intangible investment), calculated as the sum of the intercept and the residual in the following regression: INTANFT = a + b*MKTRF + c*SMB +d*HML+e*RMW+ f *CMA +$e$. Variable definitions are provided in Appendix A. We report the intercepts, slopes for HML, RMW, and INTANFT_Org, and *t*-statistics for these coefficients.

| INTAN | Low | 2 | 3 | High | Low | 2 | 3 | High | Low | 2 | 3 | High | Low | 2 | 3 | High |
|-------|-----|---|---|------|-----|---|---|------|-----|---|---|------|-----|---|---|------|
| | Panel A: MTB-INTAN Portfolios | | | | | | | | Panel B: OP-INTAN Portfolios | | | | | | | |

| | $\alpha$ | | | | $t(\alpha)$ | | | | $\alpha$ | | | | $t(\alpha)$ | | |
|---|---|---|---|---|---|---|---|---|---|---|---|---|---|---|---|---|
| Low | 0.006 | 0.005 | 0.007 | 0.007 | 3.43 | 3.24 | 4.01 | 2.85 | 0.000 | 0.001 | 0.001 | -0.001 | 0.34 | 0.25 | 0.44 | -0.43 |
| 2 | 0.004 | 0.005 | 0.003 | 0.005 | 4.89 | 4.15 | 2.28 | 2.27 | 0.004 | 0.003 | -0.001 | 0.002 | 2.44 | 1.95 | -0.36 | 1.18 |
| 3 | 0.004 | 0.004 | 0.000 | 0.002 | 6.24 | 4.63 | 0.42 | 1.27 | 0.003 | 0.004 | 0.002 | 0.001 | 2.95 | 4.56 | 1.95 | 1.30 |
| 4 | 0.002 | 0.002 | -0.001 | 0.000 | 3.49 | 2.09 | -0.68 | 0.17 | 0.002 | 0.004 | 0.001 | 0.003 | 3.14 | 4.22 | 2.09 | 2.78 |
| High | 0.001 | -0.003 | -0.002 | -0.002 | 0.97 | -2.53 | -3.12 | -1.90 | 0.003 | 0.003 | 0.002 | 0.002 | 3.29 | 3.02 | 1.89 | 1.54 |
| | $b_3$ | | | | $t(b_3)$ | | | | $b_3$ | | | | $t(b_3)$ | | | |
| Low | 0.348 | 0.363 | -0.020 | -0.401 | 5.36 | 5.77 | -0.28 | -3.90 | 0.066 | 0.162 | -0.371 | -0.604 | 0.63 | 1.43 | -3.81 | -6.80 |
| 2 | 0.448 | 0.361 | 0.024 | -0.303 | 13.46 | 8.30 | 0.45 | -3.49 | 0.025 | 0.203 | -0.267 | -0.427 | 0.42 | 3.43 | -4.65 | -8.41 |
| 3 | 0.322 | 0.393 | 0.028 | -0.276 | 13.74 | 10.93 | 0.69 | -4.36 | 0.223 | 0.379 | -0.012 | -0.282 | 5.63 | 10.61 | -0.34 | -6.86 |
| 4 | 0.200 | 0.223 | -0.076 | -0.485 | 8.20 | 5.98 | -2.29 | -9.69 | 0.282 | 0.331 | 0.038 | -0.218 | 8.97 | 9.67 | 1.38 | -5.77 |
| High | -0.020 | 0.003 | -0.317 | -0.604 | -0.52 | 0.07 | -11.32 | -13.57 | 0.184 | 0.233 | 0.042 | -0.301 | 4.49 | 5.09 | 1.29 | -4.74 |
| | $b_4$ | | | | $t(b_4)$ | | | | $b_4$ | | | | $t(b_4)$ | | | |
| Low | -0.151 | -0.044 | -0.210 | -0.780 | -2.21 | -0.66 | -2.79 | -7.18 | -0.619 | -0.414 | -0.856 | -1.183 | -5.65 | -3.47 | -8.32 | -12.60 |
| 2 | 0.071 | 0.058 | -0.113 | -0.442 | 2.03 | 1.25 | -2.03 | -4.82 | -0.204 | -0.205 | -0.497 | -0.696 | -3.23 | -3.28 | -8.20 | -12.96 |
| 3 | 0.216 | 0.225 | -0.010 | -0.519 | 8.73 | 5.91 | -0.23 | -7.76 | 0.185 | 0.126 | -0.079 | -0.289 | 4.41 | 3.30 | -2.10 | -6.67 |
| 4 | 0.305 | 0.326 | -0.048 | -0.544 | 11.81 | 8.28 | -1.37 | -10.30 | 0.370 | 0.355 | 0.245 | 0.019 | 11.13 | 9.80 | 8.54 | 0.47 |
| High | 0.107 | 0.122 | -0.303 | -0.744 | 2.58 | 2.46 | -10.21 | -15.79 | 0.346 | 0.349 | 0.242 | -0.288 | 8.00 | 7.21 | 7.08 | -4.30 |
| | $b_7$ | | | | $t(b_7)$ | | | | $b_7$ | | | | $t(b_7)$ | | | |
| Low | -0.079 | -0.004 | 0.532 | 1.332 | -0.96 | -0.06 | 5.89 | 10.22 | 0.013 | 0.041 | 0.944 | 1.404 | 0.10 | 0.28 | 7.65 | 12.47 |
| 2 | -0.212 | -0.079 | 0.472 | 1.118 | -5.04 | -1.43 | 7.10 | 10.17 | -0.015 | -0.060 | 0.679 | 0.910 | -0.20 | -0.80 | 9.34 | 14.13 |
| 3 | -0.293 | -0.219 | 0.454 | 1.010 | -9.88 | -4.81 | 8.75 | 12.60 | -0.117 | -0.132 | 0.258 | 0.584 | -2.33 | -2.92 | 5.77 | 11.23 |
| 4 | -0.352 | -0.262 | 0.347 | 0.915 | -11.38 | -5.54 | 8.25 | 14.44 | -0.250 | -0.256 | 0.183 | 0.392 | -6.27 | -5.89 | 5.32 | 8.17 |
| High | -0.275 | -0.108 | 0.324 | 0.779 | -5.53 | -1.82 | 9.13 | 13.79 | -0.151 | -0.115 | 0.113 | 0.697 | -2.91 | -1.98 | 2.75 | 8.67 |

**Table 8**

**Regressing monthly excess returns on components of profitability factor by OP sorting**

This table reports the regression results of monthly excess returns, R(t) – RF(t), on a set of factors over the period 1993–2022, 360 months. The model is as follows:

$$R(t) - RF(t) = \alpha + b_1 * MKTRF(t) + b_2 * SMB(t) + b_3 * HML(t) + b_4 * RMW\_Org(t) + b_5 * RMW\_INTAN(t) + b_6 * CMA(t) + b_7 * UMD(t) + e(t)$$

Firms are divided into five portfolios based on their operating profitability (OP) at the end of the previous calendar year, defined as revenue minus cost of sales, minus SG&A expenses, minus interest expense, divided by book equity. R(t) is the average monthly returns on all the stocks in a portfolio. RF(t) is the one-month Treasury bill rate observed at the beginning of the month. MKTRF (market return minus one-month Treasury bill rate) is the market factor; SMB (small minus big) is the size factor; HML (high minus low B/M) is the value factor; CMA (conservative minus aggressive investment) is the investment factor; and UMD (up minus down return) is the momentum factor. RMW_Org is orthogonal version of profitability factor RMW (robust minus weak OP), calculated as the sum of the intercept and the residual in the following regression run for the period 1993–2022: RMW=a+b*INTANFT+e, while RMW_INTAN is the intangible investment component of profitability factor RMW, calculated as the predicted RMW of the regression. Variable definitions are provided in Appendix A. The *t*-values are presented in parenthesis. *, **, and *** denote statistical significance at the 10%, 5%, and 1% levels, respectively.

| | intercept | MKTRF | SMB | HML | RMW_Org | RMW_INTAN | CMA | UMD | $R^2$ |
|---|---|---|---|---|---|---|---|---|---|
| Low OP | -0.001 | 0.902*** | 0.682*** | 0.217** | -0.399*** | -2.200*** | 0.065 | -0.355*** | 0.852 |
| | (-0.25) | (17.88) | (9.34) | (2.04) | (-3.68) | (-13.23) | (0.55) | (-8.44) | |
| 2 | 0.001 | 0.852*** | 0.606*** | 0.114* | -0.174*** | -1.020*** | 0.125 | -0.250*** | 0.907 |
| | (1.14) | (30.01) | (14.74) | (1.90) | (-2.85) | (-10.90) | (1.87) | (-10.57) | |
| 3 | 0.002*** | 0.924*** | 0.701*** | 0.170*** | 0.089** | -0.123** | 0.042 | -0.153*** | 0.946 |
| | (3.15) | (49.53) | (25.93) | (4.32) | (2.21) | (-2.00) | (0.96) | (-9.88) | |
| 4 | 0.002*** | 0.984*** | 0.664*** | 0.120*** | 0.260*** | 0.339*** | 0.013 | -0.141*** | 0.962 |
| | (4.05) | (66.11) | (30.80) | (3.84) | (8.14) | (6.91) | (0.37) | (-11.32) | |
| High OP | 0.002*** | 1.040*** | 0.623*** | 0.106** | 0.248*** | 0.060 | 0.015 | -0.164*** | 0.937 |
| | (2.70) | (49.71) | (20.58) | (2.39) | (5.52) | (0.87) | (0.30) | (-9.40) | |

**Table 9**

**Regressing monthly excess returns on orthogonal version of intangible intensity factor by analysts' expected growth in EPS (LTG)**

This table reports the regression results of monthly excess returns, R(t) – RF(t), on a set of factors over the period 1993–2022, 360 months. The model is as follows:

$$R(t) - RF(t) = \alpha + b_1 * MKTRF(t) + b_2 * SMB(t) + b_3 * HML(t) + b_4 * RMW(t) + b_5 * CMA(t) + b_6 * UMD(t) + b_7 * INTANFT\_Org(t) + e(t)$$

Panel A reports the regression results for the entire sample period 1993–2022. Panel B reports the regression results for the Internet bubble period 1995–2000. Panel C reports the regression results for the non-Internet bubble period: 1993–1994 and 2001–2022. On December of previous year between 1993 and 2022, firms are divided into five portfolios based on analysts expected long-term growth in earnings per share (LTG). R(t) is the average monthly returns on all the stocks in a portfolio. RF(t) is the one-month Treasury bill rate observed at the beginning of the month. MKTRF (market return minus one-month Treasury bill rate) is the market factor; SMB (small minus big) is the size factor; HML (high minus low B/M) is the value factor; RMW(robust minus weak profitability) is the profitability factor; CMA (conservative minus aggressive investment) is the investment factor; UMD (up minus down return) is the momentum factor; and INTANFT_Org is orthogonal version of intangible intensity factor INTANFT (high minus low intangible investment), calculated as the sum of the intercept and the residual in the following regression run for the period 1993–2022: INTANFT = a + b*MKTRF + c*SMB + d*HML + e*RMW + f*CMA + e. Variable definitions are provided in Appendix A. The *t*-values are presented in the parenthesis. *, **, and *** denote statistical significance at the 10%, 5%, and 1% levels, respectively.

Panel A: Regressions for the entire sample period 1993–2022

| | intercept | MKTRF | SMB | HML | RMW | CMA | UMD | INTANFT_Org | $R^2$ |
|---|---|---|---|---|---|---|---|---|---|
| Low LTG | -0.001 | 0.978*** | 0.332*** | 0.285*** | 0.395*** | 0.094** | -0.132*** | -0.206*** | 0.937 |
| | (-0.81) | (56.52) | (13.60) | (9.88) | (12.94) | (2.24) | (-8.91) | (-5.63) | |
| 2 | 0.002** | 0.942*** | 0.436*** | 0.141*** | 0.276*** | 0.088** | -0.110*** | -0.132*** | 0.949 |
| | (2.41) | (62.40) | (20.49) | (5.62) | (10.37) | (2.41) | (-8.51) | (-4.15) | |
| 3 | 0.002*** | 0.991*** | 0.552*** | 0.072** | 0.182*** | 0.018 | -0.136*** | -0.020 | 0.946 |
| | (3.11) | (57.50) | (22.66) | (2.51) | (6.00) | (0.42) | (-9.23) | (-0.56) | |
| 4 | 0.003*** | 1.073*** | 0.715*** | -0.067** | -0.047 | -0.031 | -0.172*** | 0.184*** | 0.964 |
| | (3.60) | (63.94) | (30.16) | (-2.40) | (-1.59) | (-0.77) | (-11.95) | (5.20) | |
| High LTG | 0.004*** | 1.137*** | 0.843*** | -0.244*** | -0.415*** | -0.072 | -0.174*** | 0.453*** | 0.963 |

|  | | | | | | | | |
|---|---|---|---|---|---|---|---|---|
| | (5.15) | (55.97) | (29.40) | (-7.21) | (-11.61) | (-1.47) | (-9.99) | (10.57) |

**Panel B: Regressions for the Internet bubble period 1995–2000**

|  | intercept | MKTRF | SMB | HML | RMW | CMA | UMD | INTANFT_Org | $R^2$ |
|---|---|---|---|---|---|---|---|---|---|
| Low LTG | 0.001 | 0.969*** | 0.292*** | 0.482*** | 0.258*** | 0.053 | -0.171*** | -0.239*** | 0.953 |
| | (0.45) | (27.19) | (8.66) | (7.80) | (4.70) | (0.70) | (-5.80) | (-4.29) | |
| 2 | -0.002 | 1.063*** | 0.517*** | 0.347*** | 0.206*** | 0.083 | -0.204*** | -0.123** | 0.954 |
| | (-1.43) | (27.04) | (13.88) | (5.08) | (3.40) | (0.99) | (-6.27) | (-2.00) | |
| 3 | -0.002 | 1.010*** | 0.657*** | 0.174* | 0.254*** | -0.157 | -0.284*** | 0.075 | 0.932 |
| | (-0.79) | (18.67) | (12.82) | (1.86) | (3.06) | (-1.37) | (-6.34) | (0.89) | |
| 4 | 0.001 | 1.152*** | 0.787*** | -0.110 | 0.072 | -0.147 | -0.317*** | 0.183** | 0.974 |
| | (0.30) | (19.88) | (14.35) | (-1.10) | (0.81) | (-1.19) | (-6.62) | (2.02) | |
| High LTG | 0.007*** | 1.082*** | 0.833*** | -0.319*** | -0.347*** | -0.274** | -0.231*** | 0.368*** | 0.972 |
| | (2.72) | (17.76) | (14.44) | (-3.02) | (-3.70) | (-2.11) | (-4.58) | (3.86) | |

**Panel C: Regressions for the non-Internet bubble period: 1993–1994 and 2001–2022**

|  | intercept | MKTRF | SMB | HML | RMW | CMA | UMD | INTANFT_Org | $R^2$ |
|---|---|---|---|---|---|---|---|---|---|
| Low LTG | -0.001 | 0.992*** | 0.381*** | 0.242*** | 0.358*** | 0.073 | -0.098*** | -0.159*** | 0.941 |
| | (-1.12) | (49.88) | (12.06) | (7.52) | (9.55) | (1.52) | (-5.37) | (-3.49) | |
| 2 | 0.002*** | 0.947*** | 0.431*** | 0.117*** | 0.220*** | 0.087** | -0.070*** | -0.049 | 0.958 |
| | (3.42) | (59.73) | (17.10) | (4.58) | (7.35) | (2.25) | (-4.83) | (-1.36) | |
| 3 | 0.003*** | 1.010*** | 0.550*** | 0.046* | 0.048 | 0.048 | -0.066*** | 0.050 | 0.964 |
| | (5.13) | (61.88) | (21.21) | (1.73) | (1.57) | (1.22) | (-4.40) | (1.34) | |
| 4 | 0.003*** | 1.062*** | 0.728*** | -0.077*** | -0.148*** | 0.034 | -0.127*** | 0.264*** | 0.974 |
| | (5.01) | (67.00) | (28.90) | (-3.02) | (-4.94) | (0.89) | (-8.73) | (7.28) | |
| High LTG | 0.004*** | 1.130*** | 0.866*** | -0.252*** | -0.451*** | -0.003 | -0.159*** | 0.504*** | 0.962 |
| | (4.80) | (50.64) | (24.43) | (-6.99) | (-10.72) | (-0.05) | (-7.78) | (9.86) | |

# Appendix A Definitions of Variables

| Variable | Definition |
|---|---|
| MTB | is market-to-book ratio obtained as the market value of equity divided by the book value of equity. |
| ROE | is obtained as net income divided by book value of equity. |
| RD | is research and development expenditures deflated by revenue. |
| SGA | is selling, general and administrative expenses deflated by revenue. |
| INTAN | is intangible capital investment, calculated as R&D expenses plus the investment portion of selling, general & administrative (SG&A) expenses, divided by average total assets. The investment portion of SG&A expenses is the difference between actual SG&A expenses and the SG&A expenses predicted by model $SG\&A_{it} = \alpha + \beta*Revenues_{it} + \gamma * Revenue\_Decrease_{it} + \lambda *Loss_{it} + e_{it}$, where SG&A and Revenues are scaled by average total assets, Revenue_Decrease is a dummy variable that equals 1 if revenue declines during the year relative to last year and 0 otherwise, and Loss is a dummy variable that equals 1 if net income during the year is negative and 0 otherwise. Regressions are performed by industry and year, based on 3-digit SIC codes. |
| R(t) – RF(t) | is the excess monthly return on a stock portfolio. R(t) is the average monthly returns on all the stocks in a portfolio. RF(t) is the one-month Treasury bill rate observed at the beginning of the month. |
| MKTRF | is the Fama-French market factor, calculated as the value-weight monthly return on all NYSE, AMEX, and NASDAQ stocks (from CRSP) minus the one-month Treasury bill rate. |
| SMB | is the Fama-French size factor, calculated as the difference between small-firms return and big-firms return. |
| HML | is the Fama-French value factor, calculated as the difference between high book-to-market equity return and low book-to-market equity return. |
| RMW | is the Fama-French profitability factor, calculated as the difference between strong operating profitability (OP) equity return and weak OP equity return, where OP is defined as revenue minus cost of sales, minus SG&A expenses, minus interest expense, divided by book equity. |
| CMA | is the Fama-French investment factor, calculated as the difference between conservative investment equity return and aggressive investment equity return, where investment is the percentage change in total assets. |
| UMD | is the momentum factor, calculated as the difference between high performing equity return and low performing equity return. |
| INTANFT | is the intangible intensity factor constructed in the following way: (1) in June of each year t over the periods 1963–1992 and 1993–2022, all NYSE, Amex and NASDAQ stocks are divided into two groups, small and big (S and B), based on the NYSE median market cap; (2) These stocks are then divided into three intangible intensity groups for the bottom (Low), middle (Medium), and top (High) of the ranked values of INTAN, defined as R&D expenses plus the investment portion of selling, general & administrative (SG&A) expenses, divided by average total assets; (3) the intangible intensity factor INTANFT is calculated as the difference, each month, between the simple average of the returns on the two high-INTAN portfolios (S/H and B/H) and the average of the returns on the two low-INTAN portfolios (S/L and B/L). |
| INTANFT_Org | is the orthogonal version of intangible intensity factor INTANFT, calculated as the sum of the intercept and the residual in the following regression run separately for the periods 1963–1992 and 1993–2022:  INTANFT = a + |

|          | b*MKTRF + c*SMB +d*HML+e*RMW+ f*CMA +$e$. |
|----------|-------------------------------------------|
| RMW_Org  | is the orthogonal version of profitability factor RMW, calculated as the sum of the intercept and the residual in the following regression run for the period 1993–2022: RMW=a+b*INTANFT+$e$. |
| RMW_INTAN | is the predicted RMW of the following regression run for the period 1993–2022: RMW=a+b*INTANFT+$e$. |